\documentstyle[pra,12pt,aps]{revtex}
\begin{document}
\draft
\author{Li-Bin Fu$^1$, Jie Liu$^1$, Shi-Gang Chen$^1$, and Yi-Shi Duan$^2$}
\address{$^1$ Institute of Applied Physics and Computational Mathematics, \\
P.O. Box 8009(28), Beijing 100088, P.R. China \\
$^2$ Institute of Theoretical Physics, Department of Physics,\\
Lanzhou University, 730000, P. R. China }
\title{The configuration of a topological current and physical structure: an
application and paradigmatic evidence}
\maketitle

\begin{abstract}

In the $\phi $-mapping theory, the topological current constructed by the
order parameters can possess different inner structure. The difference in
topology must correspond to the difference in physical structure. The
transition between different structures happens at the bifurcation point of
the topological current. In a self-interaction two-level system, the change
of topological particles corresponds to change of energy levels. 
\end{abstract}


\vskip 0.5cm In recent years, topology has established itself as
an important part of the physicist's mathematical arsenal
\cite{zh1}. The concepts of the topological particle and its
current have been widely used in particle physics \cite{duan1,hha}
and topological defect theory \cite {zh4}. Here, the topological
particles are regarded as abstract particles, such as monopoles
and the points defects.

In this paper, we give a new understanding in topology and
physics. Many physics system can be described by employing the
order parameters. By making use of the $\phi $-mapping theory, we
find that the topological current constructed by the order
parameters can possess different inner structure. The topological
properties are basic properties for a physics system, so the
difference in configuration of the topological current must
correspond to the difference in physical structure.

Considering a $(n+1)$-dimensional system with $n$-component vector order
parameter field $\vec \phi ({\bf x}),$ where ${\bf x}=(x^0,x^1,x^2,\cdots
x^n)$ correspond to local coordinates. The direction unit field of $\vec \phi
$ is defined by
\begin{equation}
n^a=\frac{\phi ^a}{||\phi ||},\quad a=1,2,\cdots n  \label{c1unit}
\end{equation}
where
\[
||\phi ||=(\phi ^a\phi ^a)^{1/2}.
\]
The topological current of this system is defined by
\begin{equation}
j^\mu (x)=\frac{\in ^{\mu \mu _1\cdots \mu _n}}{A(S^{n-1})(d-1)!}\in
_{a_1\cdots a_n}\partial _{\mu _1}n^{a_1}\cdots \partial _{\mu _n}n^{a_n}
\label{firstcurr}
\end{equation}
where $A(S^{n-1})$ is the surface area of $(n-1)$-dimensional unit sphere $%
S^{n-1}$. Obviously, the current is identically conserved,
\[
\partial _\mu j^\mu =0.
\]
If we define a Jacobians by
\begin{equation}
\in ^{a_1\cdots a_n}D^\mu (\frac \phi x)=\in ^{\mu \mu _1\cdots \mu
_n}\partial _{\mu _1}\phi ^{a_1}\cdots \partial _{\mu _n}\phi ^{a_n},
\label{firstjac}
\end{equation}
as has been proved before \cite{topc}, this current takes the form as
\begin{equation}
j^\mu =\delta (\vec \phi )D^\mu (\frac \phi x).  \label{deltfirstcurr}
\end{equation}
Then, we can obtain
\begin{equation}
j^\mu =\sum_{i=1}^l\beta _i\eta _i\delta (\vec x-\vec z_i(x^0))\frac{%
dz_i^\mu }{dx^0},
\end{equation}
in which $z_i(x^0)$ are the zero lines where $\vec \phi ({\bf x)}=0,$ the
positive integer $\beta _i$ and $\eta _i=sgnD(\frac \phi {\vec x})$ are the
Hopf index and Brouwer degree of $\phi $-mapping \cite{zh17} respectively,
and $l$ is the total number of the zero lines$.$ This current is similar to
a current of point particles and the $i$-th one with the charge $\beta
_i\eta _i,$ and the zero lines $z_i(x)$ are just the trajectories of the
particles, for convenience we define these point particles as topological
particles. Then the total topological charge of the system is
\[
Q=\int_Mj^0d^nx=\sum_{i=1}^l\beta _i\eta _i,
\]
here $M$ is a $n$-dimensional spatial space for a given $x^0.$ This is a
topological invariant and corresponds to some basic conditions of this
physical system. However, it is important that the inner structure of the
topological invariant can be constructed in different configurations, i.e.,
the number of the topological particles and their charge can be changed.
This change in configuration of the topological current must correspond to
some change in physical structure.

All of the above discussion is based on the condition that
\[
D\left( \frac \phi x\right) =\left. D^0\left( \frac \phi x\right) \right|
_{z_i}\neq 0.
\]
When $\left. D\left( \frac \phi x\right) \right| _{z_i}=0$ at some points $%
p_i^{*}=z^{*}(x_c^0)$ at $x^0=x_c^0$ along the zero line $z_i(x^0)$, it is
shown that there exist several crucial cases of branch process, which
correspond to the topological particle generating or annihilating at limit
points and splitting, encountering or merging at the bifurcation points. A
vast amount of literature has been devoted to discussing these features of
the evolution of the topological particles \cite{bifur}. Here, we will not
spend more attention on describing these evolution, but put our attention on
the physical substance of these processes.

As we have known before, all of these branch processes keep the total
topological charge conserved, but it is very important that these branch
processes change the number and the charge of the topological particles.
i.e. change the inner structure of the topological current. In our point of
view, the different configuration of topological current corresponds to the
different physical structure.

We consider $x^0$ as a parameter $\lambda $ of a physics system. Let us
define
\begin{equation}
\left. f_i(\lambda )=D^0\left( \frac \phi x\right) \right| _{z_i}
\end{equation}
As $\lambda $ changing, the value of $f_i(\lambda )$ changes along the zero
lines $z_i(\lambda ).$ At a critical point $\lambda =\lambda _c,$ when $%
f_i(\lambda _c)=0,$ we know that the inner structure of the topological
current will be changed in some way, at the same time the physical structure
will also be changed, i.e., the physical structure when $\lambda <\lambda _c$
will be different from the one when $\lambda >\lambda _c.$ The transition
between these structures occurs at the bifurcation points where $f_i(\lambda
)=0.$

As an application and example, let us consider a self-interacting two-level
model introduced in Ref. \cite{wuniu}. The nonlinear two-level system is
described by the dimensionless Schr\"odinger equation
\begin{equation}
i\frac \partial {\partial t}\left(
\begin{array}{c}
a \\
b
\end{array}
\right) =H(\gamma )\left(
\begin{array}{c}
a \\
b
\end{array}
\right)  \label{a}
\end{equation}
with the Hamiltonian given by
\begin{equation}
H(\gamma )=\left(
\begin{array}{cc}
\frac \gamma 2+\frac C2(|b|^2-|a|^2) & \frac V2 \\
\frac V2 & -\frac \gamma 2-\frac C2(|b|^2-|a|^2)
\end{array}
\right) ,  \label{b}
\end{equation}
in which $\gamma $ is the level separation, $V$ is the coupling constant
between the two levels, and $C$ is the nonlinear parameter describing the
interaction. The total probability $|a|^2+|b|^2$ is conserved and is set to
be $1$.

We assume $a=|a|e^{i\varphi _1(t)}$, $b=|b|e^{i\varphi _2(t)},$ the
fractional population imbalance and relative phase can be defined by
\begin{equation}
z(t)=|b|^2-|a|^2,\;\;\;\;\;\varphi (t)=\varphi _2(t)-\varphi _1(t).
\label{bal}
\end{equation}
From Eqs. (\ref{a}) and (\ref{b}), we obtain
\begin{equation}
\frac d{dt}z(t)=-V\sqrt{1-z^2(t)}\sin [\varphi (t)]  \label{ceq1}
\end{equation}
\begin{equation}
\frac d{dt}\varphi (t)=\gamma +Cz(t)+\frac{Vz(t)}{\sqrt{1-z^2(t)}}\cos
[\varphi (t)].  \label{ceq2}
\end{equation}

If we chose $x=2|a||b|\cos (\varphi ),$ $y=2|a||b|\sin (\varphi )$, it is
easy to see that $x^2+y^2+z^2=1$ by considering $|a|^2+|b|^2=1,$ which
describes a unit sphere $S^2\ $with $z$ and $\varphi $ a pair of co-ordinates%
$.$ We define a vector field on this unit sphere:
\begin{equation}
\phi ^1=-V\sqrt{1-z^2}\sin (\varphi ),  \label{f1}
\end{equation}
\begin{equation}
\phi ^2=\gamma +Cz+\frac{Vz}{\sqrt{1-z^2}}\cos (\varphi )  \label{f2}
\end{equation}
Apparently, there are singularities at the two pole points $z=\pm
1,$ which make the vector $\vec \phi $ is discontinuous at these
points. However, the direction unit vector $\vec n$ is continuous.
In the $\phi $-mapping theory, we only need the unit vector $\vec
n$ is continuous and differentiable on the whole sphere $S^2$ (at
the zero points of $\vec \phi $, the differential of $\vec n$ is a
general function)$,$ and the vector $\vec \phi $ is continuous and
differentiable at the neighborhoods of its zero points. Then from $\phi $%
-mapping theory, we can obtain a topological current as
\begin{equation}
\vec j=\sum_{i=1}^l\beta _i\eta _i\delta (\varphi -\varphi _i(\gamma
))\delta (z-z_i(\gamma ))\frac{dz_i}{d\gamma }\left| _{p_i}\right. ,
\end{equation}
in which $p_i=p_i(z_i,\varphi _i)$ is the trajectory of the $i$-th
topological particle $P_i$ and
\begin{equation}
\eta _i=sgn\left( D(\gamma )\right) =sgn\left( \det \left. \left(
\begin{array}{cc}
\partial \phi ^1/\partial \varphi  & \partial \phi ^2/\partial \varphi  \\
\partial \phi ^1/\partial z & \partial \phi ^2/\partial z
\end{array}
\right) \right| _{p_i}\right) .
\end{equation}
From Eqs. (\ref{f1}) and (\ref{f2}), it is easy to see that $\vec \phi $ is
single-valued on $S^2,$ which states that Hopf index $\beta _i=1$ $%
(i=1,2,\cdots ,l)$ here. It can be proved that the total charge of this
system is just the Euler number of $S^2$ which is $2$ \cite{lishen}. This is
a topological invariant of $S^2$ which corresponds to the basic condition of
this system: $|a|^2+|b|^2=1.$ The following discussion can show that this
topological invariant can possess different configuration when $\gamma $
changes. This difference in topology corresponds to the change in adiabatic
levels of this nonlinear system.

We can prove that every topological particle corresponds to an eigenstate of
the nonlinear two-level system. By solving $\vec \phi =0$ form (\ref{f1})
and (\ref{f2}), we find there are two different cases for discussing.

Case 1. For $|C/V|\leq 1,$ there only two topological particles $P_1$ and $%
P_2$, which locate on the line $\varphi =0$ and $\varphi =\pi $ respectively
as shown in the upper panel of Fig.1. All of them with topological charge $%
+1,$ and $D(\gamma )|_{P_{1,2}}>0$ for any $\gamma .$ Correspondingly, in
this case, there are only two adiabatic energy levels in this nonlinear
two-level mode for various $\gamma $ \cite{wuniu}, as shown in the lower
panel of Fig.1, $P_1$ corresponds to the upper level and $P_2$ corresponds
to the lower level.

Case 2, For $C/V>1,$ two more topological particles can appear when $\gamma $
lies in a window $-\gamma _c<\gamma <\gamma _c$. The boundary of the window
can be obtained by assuming $D(\gamma )|_{P_i}=0,$ yielding
\begin{equation}
\gamma _c=(C^{2/3}-V^{2/3})^{3/2}.  \label{gc}
\end{equation}
The striking feature happens at $\gamma =-\gamma _c,$ there exists a
critical point $T_1^{*}(z_c,\pi )$ with $D(\gamma _c)|_{T_1^{*}}=0,$ as we
have shown in Ref. \cite{bifur}, we can prove that this point is a limit
point, and a pair of topological particles $P_3$ and $P_4$ generating with
opposite charge $-1$ and $+1$ respectively$,$ both of the new topological
particles lie on the line $\varphi =\pi $. One of the original topological
particle, $P_2$ with charge $+1$ on the line $\varphi =\pi ,$ moves smoothly
up to $\gamma =\gamma _c,$ where it collides with $P_3$ and annihilates with
it at another limit point $T_2^{*}(-z_c,\pi ).$ The other, $P_1,$ which lies
on the line $\varphi =0$, still moves safely with $\gamma .$

As pointed out by B. Wu and Q. Niu \cite{wuniu}, when the interaction is
strong enough $(C/V>1),$ a loop appears at the tip of the lower adiabatic
level when $C/V>1$ while $-\gamma _c\leq \gamma \leq \gamma _c$. We show the
interesting structure in Fig. 2 in which $C/V=2$. For $\gamma <-\gamma _c,$
there are two adiabatic levels, the upper level corresponds to the
topological particle $P_1$, the lower one corresponds to the topological
particle $P_2;$ for $\gamma >\gamma _c$, there are also only two adiabatical
levels, but at this time the lower one corresponds to $P_4$ while the upper
one still corresponds to $P_1.$ The arc part of the loop on the tip of lower
level when $-\gamma _c<\gamma <\gamma _c$ just corresponds to $P_3,$ which
merges with the level corresponding to $P_4$ at the point $M$ on the left
and with the one corresponding to $P_1$ at the point $T$ on the right$.$

From the above discussion, one finds that when the structure of the
topological current is changed by generating or annihilating a pair of
topological particles ( the upper panel of Fig. 2), at the same time, the
physical structure is changed by adding two energy levels or subtracting two
energy levels respectively (the lower panel of Fig.2). The critical
behaviors happen at the limit points where $D(\gamma _c)|_{T_{1,2}^{*}}=0$ .

In fact, this nonlinear two-level model is of comprehensive interest for it
associates with a wide range of concrete physical systems, e.g., BEC in an
optical lattice\cite{becol1} or in a double-well potential\cite{dwpp1}, and
the motion of small polarons\cite{polar}. So, the relation between
topological particles and physical inner structure can be observed in
experimental methods. Here we propose a perspective system for observing the
striking phenomenon: a Bose-Einstein condensate in a double-well potential%
\cite{dwpp1,RS99}. The amplitudes of general occupations $N_{1,2}(t)$ and
phases $\varphi _{1,2}$ obey the nonlinear two-mode Schr\"odinger equations,
approximately\cite{RS99},
\begin{eqnarray}
i\hbar \frac{\partial \phi _1}{\partial t} &=&(E_1^0+U_1N_1)\phi _1-K\phi _2
\nonumber \\
i\hbar \frac{\partial \phi _2}{\partial t} &=&(E_2^0+U_2N_2)\phi _2-K\phi _1
\end{eqnarray}
with $\phi _{1,2}=\sqrt{N_{1,2}}exp(i\varphi _{1,2})$, and total number of
atoms $N_1+N_2=N_T$, is conserved. Here $E_{1,2}^0$ are zero-point energy in
each well, $U_{1,2}N_{1,2}$ are proportional to the atomic self-interaction
energy, and $K$ describes the amplitude of the tunnelling between the
condensates. After introducing the new variables $z(t)=(N_2(t)-N_1(t))/N_T$
and $\varphi =\varphi _2-\varphi _1$, one also obtain an equations having
the same form as Eqs. (\ref{a}) and (\ref{b}) except for the parameters
replaced by
\begin{equation}
\gamma =-[(E_1^0-E_2^0)-(U_1-U_2)N_T/2]/\hbar ,
\end{equation}
\begin{equation}
V=2K/\hbar ,\;C=(U_1+U_2)N_T/2\hbar .
\end{equation}
With these explicit expressions, our theory and results can be directly
applied to this system without intrinsic difficulty. In this system the
topological particles can be located by the stable occupation and relative
phase $\varphi =0$ or $\varphi =\pi $ for a give parameter $\gamma .$ And,
one can draw the zero line of each topological particle by giving different $%
\gamma .$ We hope our discussions will stimulate the experimental works in
the direction.

We note that for a system the global property (topology) is given, the
interesting feature is that under the same topology the topological
configuration can be different, this difference must correspond the
different physical structure. This relation between topological
configuration and physical structure gives an important property to classify
some physical system which contains many different structures.

We thank Prof. B.Y. Ou for useful discussions. This project was supported by
Fundamental Research Project of China.

\section{Figure caption}

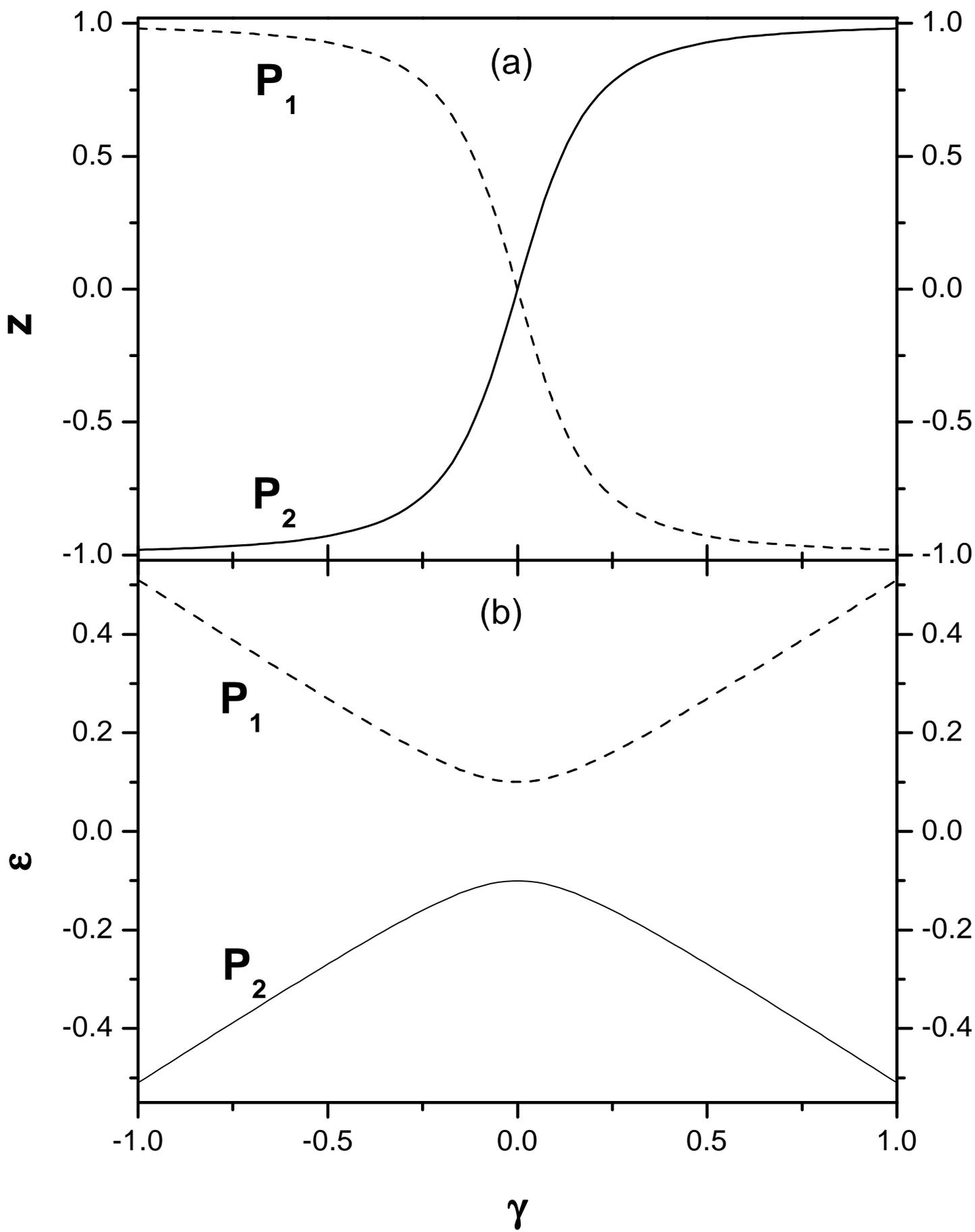
Fig. 1. (a) The projecting of trajectory of topological particles on $%
(z-\gamma )$ plane for $C/V=0.$ $P_i$ denotes the $i$-th topological
particle. (b). The energy levels for $C/V=0.$ Each level is labelled by the
topological particle which corresponds to it.

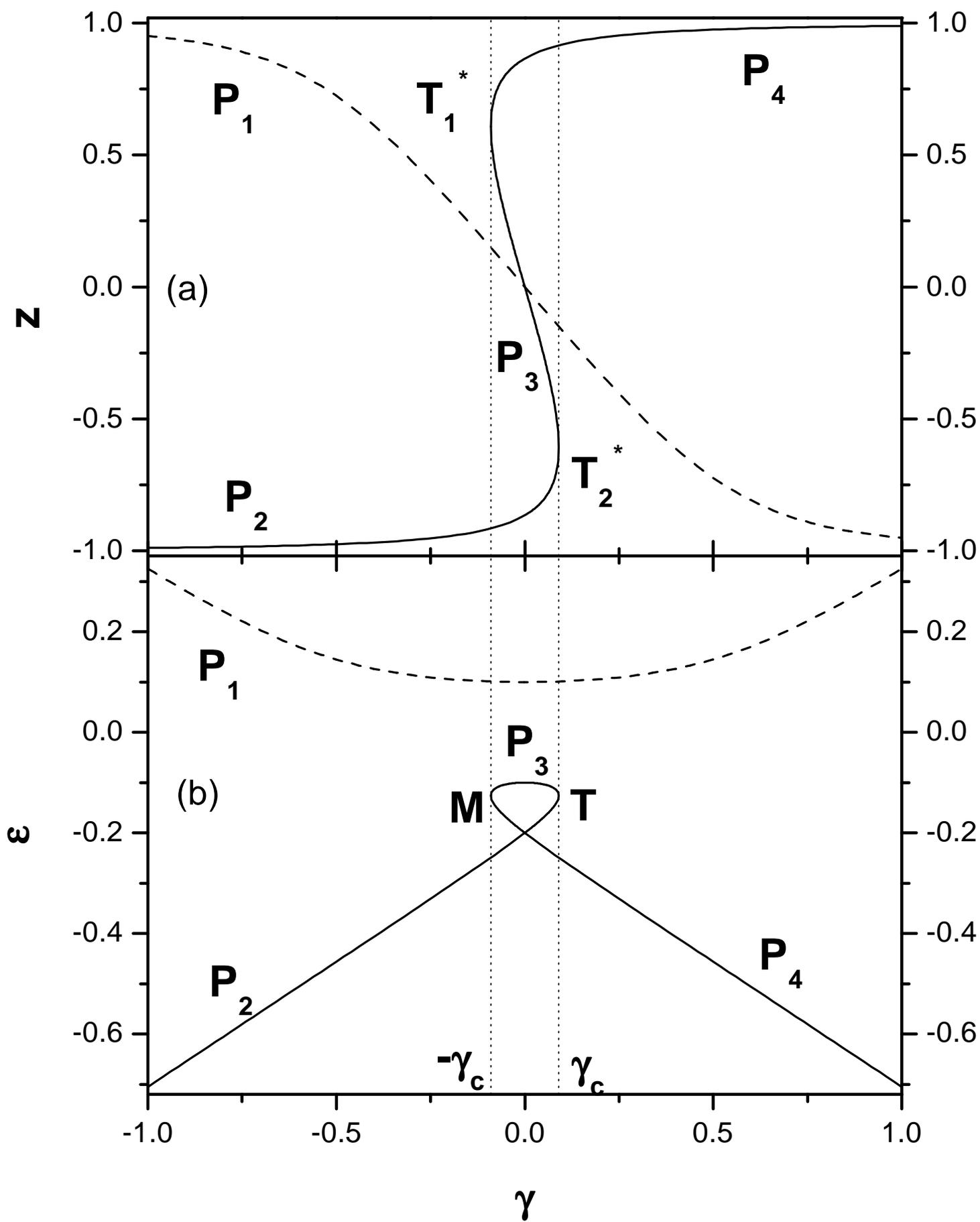
Fig. 2. (a) The projecting of trajectory of topological particles on $%
(z-\gamma )$ plane for $C/V=2.$ $P_i$ denotes the $i$-th topological
particle. (b). The energy levels for $C/V=2.$ Each level is labelled by the
topological particle which corresponds to it.

\end{document}